\documentclass[aps,pra,preprint,amsmath,amssymb,superscriptaddress,nofootinbib]{revtex4-1}
\usepackage{graphicx,wrapfig,float}

\begin{document}


\title{Spectroscopy of ultracold neutrons using diffraction by a moving grating}


\author{G.V. Kulin}\email[kulin@nf.jinr.ru] \\
\author{A.I. Frank}
\author{S.V. Goryunov}
\affiliation{Joint Institute for Nuclear Research, Dubna, Russia}
\author{P. Geltenbort}
\author{M. Jentschel}
\affiliation{Institut Lauer-Langevin, Grenoble, France}
\author{V.A. Bushuev}
\affiliation{Moscow State University, Russia}
\author{B. Lauss}
\author{Ph. Schmidt-Wellenburg}
\affiliation{Paul Scherrer Institut, Switzerland}
\author{A. Panzarella}
\author{Y. Fuchs}
\affiliation{ESRF,Grenoble, France}

\date{24 December 2015}

\begin{abstract}
Spectra of ultracold neutrons that appeared in experiments on neutron diffraction by a moving grating were measured using the time-of-flight Fourier spectrometer. Diffraction lines of five orders were observed simultaneously. The obtained data are in good agreement with the theoretical predictions based on the multiwave dynamical theory of neutron diffraction by a moving grating.
\end{abstract}

\pacs{}

\maketitle

\section{Introduction}
It is known that the neutron is a very convenient object for the demonstration and investigation of the wave properties of massive particles. In the majority of performed or proposed experiments a neutron beam (or a wave) is constant, and experimental results may be described by the stationary Schrödinger equation. But neutron optics acquires essentially new qualities when any parameter describing the interaction of the neutron wave with an object varies with time. The nonstationary action on the neutron wave makes it possible to significantly change its properties, such as energy spectrum, spin, intensity, phase, direction of propagation, and so on. Moshinsky \cite{MoshinskyPhysRev881952} was apparently the first to discuss the problem of nonstationary quantum effects in the optics of massive particles. He considered the evolution of the wave function after the instantaneous removal of a perfect absorber from a beam of monochromatic particles. His result for the wave evolution in the right half-space coincided in form with the familiar pattern of the Fresnel diffraction of light at an abrupt edge. That is why he named the considered phenomenon "diffraction in time". Gerasimov and Kazarnovsky \cite{GerasimovSovPhysJETP441976} considered the possibility of the observation of a number of nonstationary quantum phenomena arising from the interaction of ultracold neutrons (UCNs) with a potential barrier oscillating in time. In a number of subsequent theoretical \cite{GahlerZPhysB561984, FelberPhysicaBC1511988, FelberPhysicaB1611990, NosovJMoscowPhysSoc11991, GolubPhysLettA1621992, NosovPhysAtomNucl571994, FrankAnnNYAcadSci7551995p293, FrankPhysAtomNucl621999, FelberFoundPhys291999} and experimental \cite{HilsPhysRevA581998} studies the evolution of the neutron flux after transmission through a fast quantum chopper was investigated in more detail.

The effect of energy quantization of polarized neutrons in the interaction with the oscillating magnetic field was observed in the experiment \cite{SummhammerPhysRevLett751995}. In Refs. \cite{FrankAnnNYAcadSci7551995p858, KozlovPhysAtomNucl582005} a number of neutron optical phenomena occurring in the reflection and refraction of neutrons at the interface of the matter with time-dependent magnetic induction were analyzed theoretically. The reflection of very cold neutrons from the vibrating surface was observed in Ref. \cite{FelberPhysRevA531996}. At a low vibration frequency this phenomenon can be considered as a classical one whereas at a high frequency and small amplitude it becomes essentially quantum.

Under certain conditions neutron diffraction may also be considered as a nonstationary phenomenon. In Ref. \cite{FrankJINRCommP488511975} UCN diffraction by surface Rayleigh waves was treated as a cause of inelastic neutron scattering resulting in a decrease in the neutron storage time in traps. This phenomenon may also be considered as a Doppler shift of the neutron wave frequency. About a decade later the diffraction of cold neutrons by surface acoustic waves generated at the surface of a quartz crystal was observed \cite{HamiltonPhysRevLett581987}. As it followed from the theoretical analysis the neutron energy shift corresponding to the first diffraction orders was about $10^{-7}$ eV relative to the initial energy. Such a change in the energy was almost three orders of magnitude less than the energy itself and was not experimentally measured. But the intensity and angular distribution of the diffracted waves were in quite satisfactory agreement with the results of the quantum calculation.

Neutron Bragg diffraction by a crystal on the surface of which the acoustic waves were excited was observed as well \cite{HamiltonPhysRevB591999}. The effect of ultrasound on neutron diffraction by perfect and mosaic crystals was studied in Refs. \cite{IolinJETP641986, MichalecPhysicaBC1511988, KuldaPhysicaBC1511988, IolinPhysicaB2412431998}. Neutron diffraction by surface waves of viscous fluids was theoretically analyzed in Refs. \cite{PokotilovskiPhysLettA2551999, LamoreauxPhysRevC662002}. This study is related to the problem of long-term storage of neutrons in traps.

Almost two decades after the publishing of the paper \cite{FrankJINRCommP488511975} the effect of neutron energy change in diffraction by a moving grating was predicted again in Ref. \cite{FrankPhysLettA1881994}. It was shown that when the amplitude or phase grating moves across the neutron beam the grating can act as a quantum modulator of the neutron wave transforming the spectrum of transmitted neutrons. As a result the spectrum is characterized by a discrete set of energies. Shortly afterwards it was proposed to use moving gratings in a neutron interferometer \cite{IoffePhysicaB2342361997} which might significantly increase its sensitivity.

The effect of quantum spectrum splitting predicted in Ref. \cite{FrankPhysLettA1881994} was observed experimentally \cite{FrankPhysLettA3112003} and investigated in more detail in Ref. \cite{FrankJETPLett812005}. An aperiodic moving grating may serve as a neutron time lens, which can be used for neutron focusing in time. This possibility was first discussed in Ref. \cite{FrankPhysAtomNucl632000}, and neutron time focusing was demonstrated experimentally later \cite{FrankJETPLett782003, BalashovPhysicaB3502004}. More recently, the possibility of deceleration and acceleration of neutrons using a blazed moving grating was discussed \cite{FrankPhysAtomNucl762013}. A smooth transition from the quantum interpretation of the problem to the classical one was demonstrated by gradually increasing the size of the grating which transforms it into a set of macroscopic lenses. It is worth noting here the theoretical study \cite{KowalskiPhysLettA3522006} also was devoted to the diffraction of neutrons and electromagnetic waves by a moving grating. In this paper the authors confined themselves only to the analysis of the possible directions of diffracted waves in the simplest case of a harmonic grating.

The nonstationary phenomenon of neutron diffraction by a moving grating has found its application in the experiments testing the weak equivalence principle for the neutron \cite{FrankJETPLett862007}. The continuation of this research \cite{FrankNIMA6112009,KulinNIMA7922015} and recent theoretical results \cite{BushuevarXivJETP2016} have brought us to the understanding of the need for more detailed investigation of UCN spectra in diffraction by a moving grating. Unfortunately, the gravity spectrometry with interference filters used in Refs. \cite{FrankPhysLettA3112003, FrankJETPLett812005, FrankJETPLett862007, FrankNIMA6112009, KulinNIMA7922015} has proved to be inadequate for the problem because it makes it possible to observe only $\pm 1$ diffraction orders.

The most appropriate method for the solution of the problem appears to be the time-of-flight (TOF) gravity spectrometry \cite{KulinJINRCommP3201472}. For this purpose the spectrometer \cite{KulinNIMA7922015} has been significantly upgraded and converted into a TOF Fourier UCN spectrometer. The spectrometer was used for new measurements of UCN spectra using diffraction by a moving phase grating. Some results of these measurements with the comparison with the theoretical prediction are reported in the present paper.

\section{Theory of neutron diffraction by a moving grating}
First of all let us show how the motion of the grating leads to the appearance of a discrete energy spectrum as reported in Refs. \cite{FrankPhysLettA1881994, FrankPhysLettA3112003, FrankJETPLett812005}. Let the wave function of the neutron incident on the grating have in the laboratory coordinate system the form
\begin{equation}
\Psi_{in}(x,z,t) = exp[i(k_{0x}x+k_{0z}z-\omega_{0} t)],
\label{eq:psiinlabframe}
\end{equation}
where $k_{0x} = MV_{0x}/\hbar$, $k_{0z} = MV_{0z}/\hbar$, $V_{0x}$ and $V_{0z}$ are the tangential and normal components of the velocity, respectively, $M$ is the neutron mass, $\hbar$ is the Planck constant,  $\omega_{0}=\hbar k^2/2M$ is the frequency and  $k_{0}=(k_{0x}^2+k_{0z}^2)^{1/2}$ is the wave number. Assuming that the grating grooves are directed along the $Y$ axis we exclude from the consideration the $y$ component of the wave vector $k_{0y}$, the presence of which does not affect the result.

\begin{figure*}[h]
		\centering
		\includegraphics[scale=1]{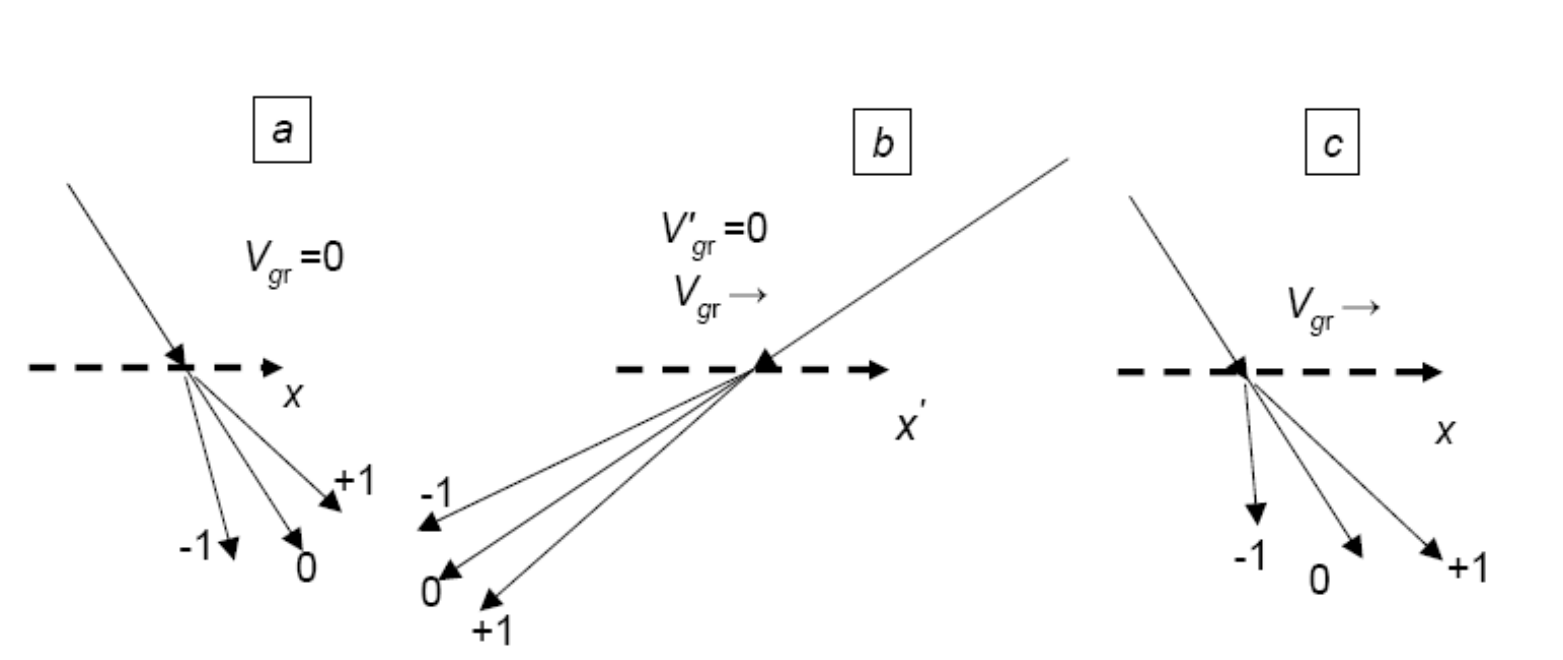}
		\caption{Schematic of neutron diffraction by a grating. (a) - Grating at rest, $\omega_{m} = \omega_{0}$, (b) - grating moving together with the coordinate system $\omega_{m} = \omega^\prime$, and (c) - grating moving in the laboratory coordinate system $\omega_{m} = \omega_{0} + m\Omega$.} 
		\vspace{1ex}
		\label{fig:diffrScheme}
\end{figure*}

Neutron diffraction by a stationary grating results in the appearance of plane waves of various orders with $x$-projections of the wave vectors $k_{mx} = k_{0x} + g_{m}$, where $g_{m} = mg_{0}$, $g_{0} = 2\pi/d$ is the value of the reciprocal lattice vector, $d$ is the spatial period of the grating, and $m = 0, \pm 1, \pm 2,\dots$, are the integer numbers [see Fig.\ref{fig:diffrScheme}(a)]. Since diffraction scattering is an elastic process, wave numbers and frequencies of all diffracted waves are equal: $k_{m} = k_{0}$, $\omega_{m} = \omega_{0}$. 

Let us assume now that the grating is moving at a constant velocity $V_{gr}$ in a positive direction of the $X$ axis. Let us choose the moving coordinate system comoving with the grating. In this system neutrons are incident on the grating at a different angle [see Fig. \ref{fig:diffrScheme}(b)] and have a wave number and energy which differ from the values in the laboratory system
\begin{equation}
\Psi^\prime_{in}(x^\prime,z,t) = A_{in}(x^\prime) exp[i(k_{0}z-\omega^\prime t)],
\label{eq:psimovframe}
\end{equation}
where
\begin{equation}
A_{in}(x^\prime)=exp(ik^\prime_{0x}x^\prime),
\label{eq:ampmovframe}
\end{equation}
$k_{0x}^\prime=k_{0x}-k_{V}$, $k_{V} = mV_{gr}/\hbar$,  $\omega^\prime= \hbar k_{0}^{\prime 2} /2M$, $k_{0}^\prime=(k_{0x}^{\prime 2}+k_{0z}^2)^{1/2}$. 

The wave function of diffracted neutrons is now
\begin{equation}
\Psi^\prime(x^\prime,z,t) =\sum_{m} a_{m}exp[i(k^\prime_{mx}x^\prime+k_{mz}(z-h)-\omega^\prime t)],
\label{eq:psidiffrmovframe}
\end{equation}
where $h$ is the thickness of the grating, $k_{mx}^\prime=k_{mx}-k_{V}$,
\begin{equation}
k_{mx}=k_{0x}+g_{m},\quad k_{0z}=(k_{0z}^2+2(k_{V}-k_{0x})g_{m}-g_{m}^2)^{1/2}.
\label{eq:diffrkvectors}
\end{equation}

For the moment we leave aside the question of the value of the amplitudes $a_{m}$ of corresponding diffraction orders.  

As in the case of the stationary grating the wave numbers $k_{m}^\prime=k_{0}^\prime$  and frequencies $\omega^\prime$ of waves of all diffraction orders are equal here, but due to the motion of the grating and the reference system they differ from $k_{0}$ and $\omega_{0}$, respectively [Fig. \ref{fig:diffrScheme}(b)]. Relative intensities of the diffraction orders are defined by the relation $I_{m} = \lvert a_{m}\rvert^2$. It should also be noted that for an arbitrary periodic phase grating, which does not absorb neutrons, the condition of flux conservation $\sum_{m} I_{m} = 1$ is satisfied. 

After returning to the laboratory system of coordinates $x = x^\prime+ V_{gr}t$ we obtain the wave function of diffracted neutrons in the form of the superposition of plain waves with amplitudes $A_{m}$, discrete frequencies $\omega_{m}$, and wave vectors  $\boldsymbol{k}_{m} = (k_{mx},k_{mz})$ [see Fig. \ref{fig:diffrScheme}(c)]
\begin{equation}
\Psi(x,z,t) =\sum_{m} A_{m}exp[i(k_{mx}x+k_{mz}z-\omega_{m} t)],
\label{eq:psidiffrlabframe}
\end{equation}
where the projections of wave vectors $k_{mx}$ and $k_{m}$ are defined by (\ref{eq:diffrkvectors}) and frequencies are $\omega_{m} = \omega_{0} + m\Omega$. The constant of spectral splitting $\Omega$ is defined by the following equivalent relations:
\begin{equation}
\Omega=2\pi/\tau=g_{0} V_{gr},
\label{eq:specsplitomega}
\end{equation}
where $\tau=d/V_{gr}$.

The relation $\sum_{m}\lvert a_{m}\rvert^2= 1$ is still valid but no longer expresses the flux conservation since the wave vectors km differ for different diffraction orders. The amplitudes $A_{m}$ in Eq. (\ref{eq:specsplitomega}) may be found from the flux conservation law
\begin{equation}
A_{m}=a_{m}[(k_{0x}^2+k_{0z}^2)/(k_{mx}^2+k_{mz}^2)]^{1/4}.
\label{eq:fluxconvlaw}
\end{equation}

In \cite{FrankPhysLettA1881994, FrankPhysLettA3112003, FrankJETPLett812005} the problem of diffraction in the moving system was analyzed in the framework of the kinematic theory when amplitudes were taken as Fourier amplitudes of the complex transmission function of an ideally flat grating. In Ref. \cite{KulinJINRCommP32004207} an attempt was made to take into account the real geometry of the phase grating by modifying its transmission function. In both cases the kinematic approach leads to the concept of complete independence of the waves with different orders $m$.

In a recent paper \cite{BushuevarXivJETP2016} the problem of neutron diffraction by a moving three-dimensional (3D) grating was analyzed in the framework of the dynamical theory when waves of different orders interact with each other as they propagate through the grating material. Below we present some results of this work. 

\begin{figure}[h]
	\centering
	\vspace{2ex}
	\includegraphics[scale=1]{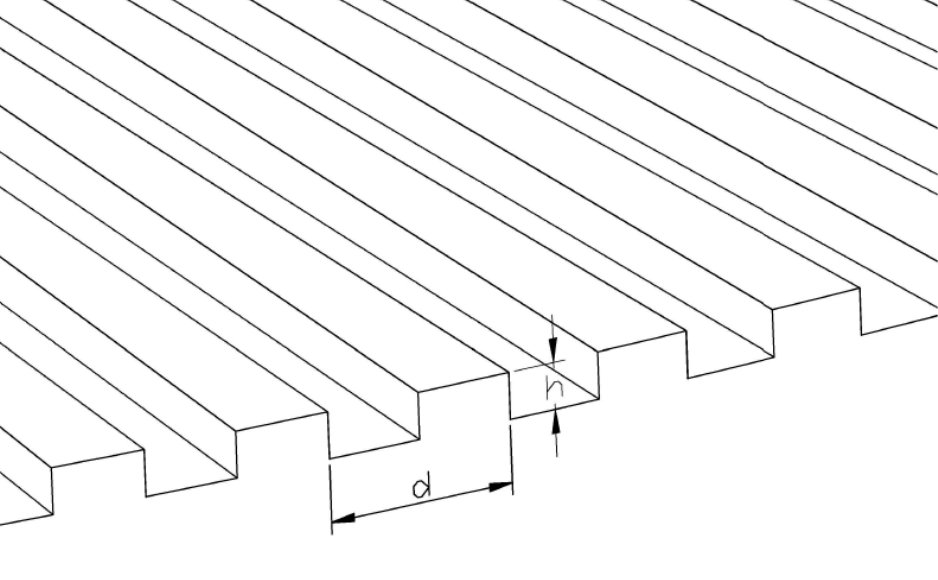}
	\caption{Profile of the phase grating.} 
	\vspace{1ex}
	\label{fig:grprofile}
\end{figure}

Let the wave with the wave function (\ref{eq:psiinlabframe}) be incident on the phase grating with period $d$ and groove depth $h$ (see Fig. \ref{fig:grprofile}) moving along the positive direction of the $X$ axis at a constant velocity $V_{gr}$. As in the above case the problem is analyzed in the moving coordinate system in which the grating is at rest. In this coordinate system we are interested only in the coordinate parts of the wave functions.

Neutron propagation in a medium is described by the Schrödinger equation,
\begin{equation}
\Delta \Psi(\boldsymbol{r})+[k^2-\chi(\boldsymbol{r})]\Psi(\boldsymbol{r})=0,
\label{eq:Shrodinger}
\end{equation}
where $k$ is the neutron wave number in a vacuum, $\chi(\boldsymbol{r}) = 4\pi N(\boldsymbol{r})b(\boldsymbol{r})$, $N(\boldsymbol{r})$ is the volume density of the nuclei, and $b(\boldsymbol{r})$ is the coherent scattering length. Let us expand the function $\chi(x)$, which is periodic in the region $0 \leq z \leq h$, into a series,
\begin{equation}
\chi(x)=\sum_{n=-\infty}^{\infty} \chi_{n} exp(ig_{n}x),
\label{eq:chiseries}
\end{equation}
where $g_{n} = ng_{0}$, $g_{0} = 2\pi/d$, $n = 0, \pm 1, \pm 2,\dots$, are integers, and 
\begin{equation}
\chi_{n}=\frac{1}{d}\int\limits_{0}^{d} \chi(x) exp(-ig_{n}x)\,dx.
\label{eq:chiintegral}
\end{equation}

A zero Fourier amplitude $\chi_{0}$ determines the value of the average grating refractive index $n_{e} = (1 - \chi_{0}/k^2)^{1/2}$ in a layer of thickness $h$. The wave function of neutrons in the region $0 \leq z \leq h$ in the moving coordinate system we can write using (\ref{eq:psimovframe}) and (\ref{eq:ampmovframe}) as the sum of the Bloch functions with amplitudes $\Psi_{m}(z)$ depending on the vertical coordinate $z$
\begin{equation}
\Psi^\prime(x^\prime,z)=\sum_{m=-\infty}^{\infty} \Psi_{m}(z) exp[i(q_{mx}x^\prime+q_{0z}z)],
\label{eq:sumBlochfuncs}
\end{equation}
where the projections of the wave vectors are
\begin{equation}
q_{mx}=k_{0x}-k_{V}+g_{m},\quad q_{0z}=(k_{0z}^2-\chi_{0}^2)^{1/2}.
\label{eq:projkvectors}
\end{equation}

Here, it was assumed that the wave number of incoming neutrons is $k^\prime = (k_{0}^2 + k_{V}^2)^{1/2}$, and the average value of $q_{0}$ in the material of the grating $0 \leq z \leq h$ is $q_{0} = k^\prime n_{e}$. 

By inserting (\ref{eq:chiseries}) and (\ref{eq:sumBlochfuncs}) into (\ref{eq:Shrodinger}) we equate the terms with equal exponents and obtain the following infinite system of coupled differential equations:
\begin{equation}
\frac{d^2 \Psi_{m}}{dz^2}+2iq_{0z} \frac{d\Psi_{m}}{dz}-\alpha_{m}\Psi_{m}-\sum_{n\neq 0} \chi_{n} \Psi_{m-n}=0,
\label{eq:syscopldiffeqs}
\end{equation}
where
\begin{equation}
\alpha_{m}=g_{m}[g_{m}-2(k_{V}-k_{0x})].
\label{eq:alpham}
\end{equation}

The solution of the system (\ref{eq:syscopldiffeqs}) in the general case is a rather difficult problem \cite{MoharamJOSA711981, GaylordApplPhysB281982, MoharamJOSA721982, MoharamJOSAA121995p1068, MoharamJOSAA121995p1077, AshkarJApplCryst432010}. It can be considerably simplified if the second derivatives are neglected. It can be shown that the condition of the smallness of their contribution is the relation $\chi_{1}\ll 4q_{0z}^2$, which is valid with the reasonable accuracy in the experiments \cite{FrankJETPLett862007, FrankNIMA6112009,KulinNIMA7922015}. As a result, we obtain the following system of reduced differential equations of the first order:
\begin{equation}
\frac{d\Psi_{m}}{dz}=-i\gamma_{m}\Psi_{m}-i\sum_{n\neq 0} \beta_{n} \Psi_{m-n},
\label{eq:sysreducdiffeqs}
\end{equation}
where $\gamma_{m} = \alpha_{m}/2q_{0z}$, $\beta_{n} = \chi_{n}/2q_{0z}$. It should be supplemented by the following boundary conditions: $\Psi_{0}(z = 0) = 1$, $\Psi_{m\neq 0}(z = 0) = 0$.

It is possible to show that the wave functions $\Psi_{m}(z)$ satisfy the flux conservation condition $\sum_{m=-\infty}^{\infty} \lvert\Psi_{m}(z)\rvert^2=1$  on any arbitrary plane $z$, whereas the flux changes from one diffraction order to another as the coordinate $z$ increases.

The coefficients $\gamma_{m}$ in (\ref{eq:sysreducdiffeqs}) and hence the functions $\Psi_{m}(z)$ depend on the grating velocity $V_{gr}$, horizontal projection of the neutron velocity $V_{0x}$, grating period $d$, and diffraction order $m$. They both vanish at $m = 0$ and Bragg condition $2k_{V} = q_{m}$.

With regard to the neutron wave function in the laboratory coordinate system in the region of observation $z > h$ it is also defined by Eqs. (\ref{eq:psidiffrlabframe}) and (\ref{eq:fluxconvlaw}) where amplitudes am should be replaced by the values of $\Psi_{m}(z = h)$ obtained from the solution of the system of Eqs. 
(\ref{eq:sysreducdiffeqs}). The condition $Re(k_{mz}) > 0$ of the undamped waves with $z$ projections of the wave vectors $k_{mz}$ defined by the Eq. (\ref{eq:diffrkvectors}) imposes certain restrictions on possible values of the grating velocity and diffraction orders. From the comparison of Eqs. (\ref{eq:diffrkvectors}) and (\ref{eq:alpham}) it follows that this condition has the form $\alpha_{m}<k_{0z}^2$.

\section{Experimental setup and measurement procedure} 
\begin{figure*}[ht]
	\centering
	\vspace{4ex}
	\minipage{0.43\linewidth}
	\includegraphics[width=\linewidth, page=1]{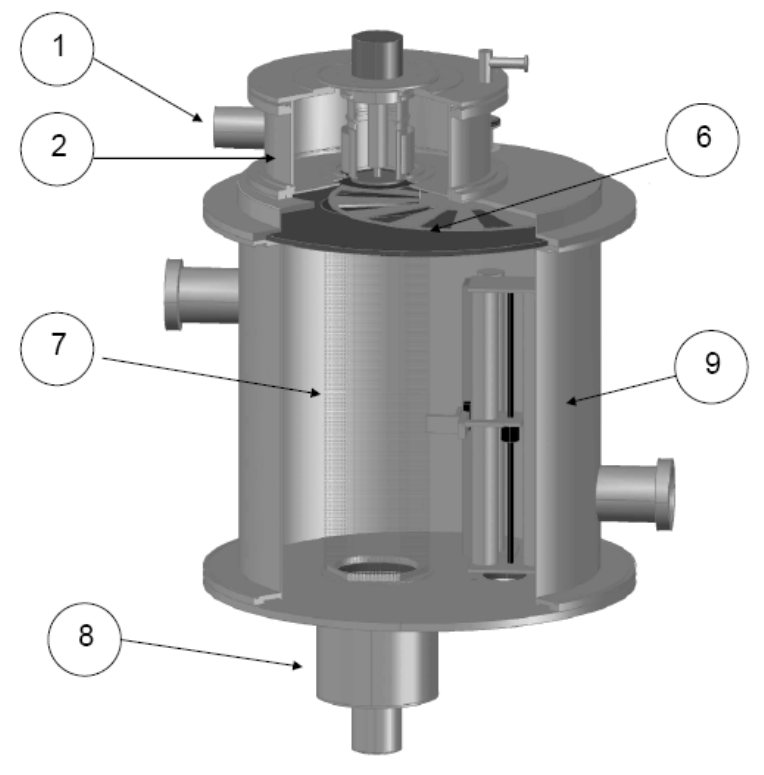}
	\endminipage\hspace{0.1\linewidth}
	\minipage{0.43\linewidth}
	\includegraphics[width=\linewidth, page=2]{figure3}
	\endminipage
	\caption{Time-of-flight Fourier spectrometer: general view (at the left) and its upper part (at the right): 1--feeding guide, 2--entrance chamber, 3--annular channel, 4--filter-monochromator, 5--grating, 6--rotor of the Fourier modulator, 7--vertical glass guide, 8--detector, 9--vacuum vessel.}
	\vspace{2ex}
	\label{fig:TOFSpectrometer}
\end{figure*}

The experiment was performed at the PF2 source of the Institute Laue-Langevin (Grenoble, France) using the time-of-flight UCN Fourier spectrometer, which is a modification of the spectrometer \cite{KulinNIMA7922015}. The detailed description of this device will be published elsewhere. The design of the spectrometer is illustrated in Fig. \ref{fig:TOFSpectrometer}. Ultracold neutrons are fed to the entrance chamber through the UCN neutron guide and, after a number of reflections, fall down the annular channel with the lower section closed by a monochromator, which is a five-layer Ni-Ti interference filter. To suppress the background of neutrons with energies higher than the effective potential of nickel, the filter monochromator is combined with a multilayer "superwindow" filter \cite{BondarenkoPhysAtomNucl621999, FrankProcSPIE37671999}. 

\begin{wrapfigure}[9]{r}{0.5\linewidth} 
	\includegraphics[width=\linewidth]{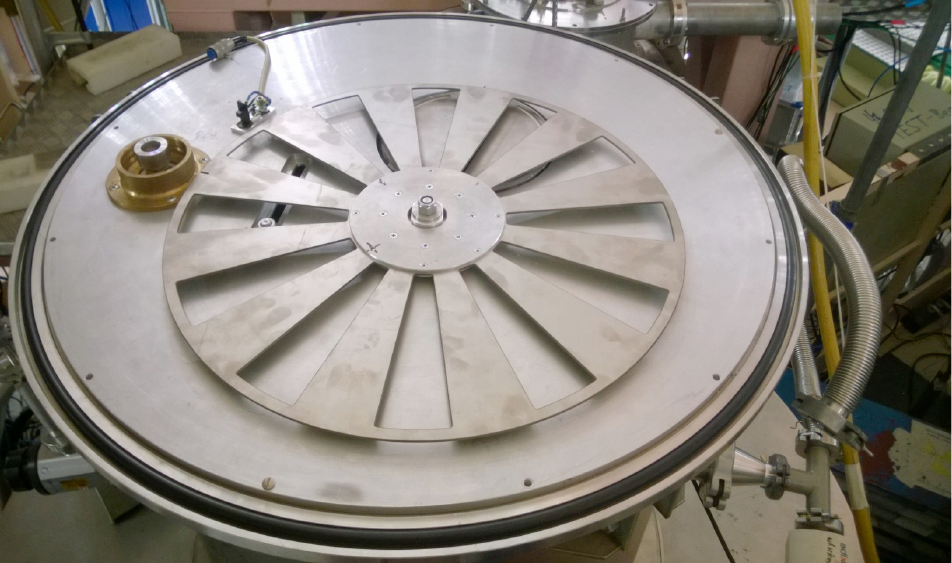}
	\caption{Rotor of the Fourier modulator}
	\label{fig:Rotor}
\end{wrapfigure}

The diffraction grating (see below) is placed just below the monochromator. The neutron flux formed by the combination of filters and transformed by the moving grating enters the spectrometric part of the device. The latter comprises a Fourier chopper, vertical guide, and detector. The Fourier chopper consists of a rotating rotor and stator. The rotor is a titanium disk about 400 mm in diameter and 2 mm thick with 12 windows in the form of sectors (see Fig. \ref{fig:Rotor}). The stator is a titanium diaphragm with only one slit. It is placed at the inlet section of the vertical neutron guide.


As in Ref. \cite{KulinNIMA7922015} the rotor is driven by a Phytron stepper motor VSS-65HV located outside the vacuum volume and connected to the rotor via a toothed belt and magnetic coupling. The rotation frequency of the rotor may reach 1800 rpm, which corresponds to a modulation frequency of 360 Hz. For the use of the electronic control system an infrared Honeywell sensor HOA2006-001 and a small slot at the periphery of the rotor are used. Stability of the rotor rotation frequency is on the order of $10^{-4}$.

UCNs passing through the modulator come to the vertical neutron guide formed by six 95 x 680mm float glass plates and reach the scintillation detector. The measurement system registered the time of arrival of pulses from the sensor of the chopper and from the detector and these data were recorded sequentially to a file. The modulation frequency $f_{k}$ ranged from 6 to 360 Hz. It was increased with a step of 6 Hz (in some measurements by 12 Hz) and upon reaching the maximum value decreased again. The duration of each measurement was usually 1000 s. The count rate was several counts per second.

The collected data were analyzed off line. The analysis included the determination of amplitude $R_{k}$ and phases $\phi_{k}$ of the count rate oscillation for each frequency $\omega_{k}=2\pi f_{k}$. The obtained set of data defined the time-of-flight spectrum in agreement with the equation
\begin{equation}
I_{exp}(t)=\frac{\pi}{2}\sum_{k} R_{k} \,sin(\omega_{k}t+\phi_{k}).
\label{eq:TOFspecseries}
\end{equation}

Due to the vertical orientation of the spectrometer and the effect of the Earth?s gravity the neutron time of flight does not linearly depend on the initial velocity. For the correct interpretation of the results it was necessary to recalculate the TOF spectrum to the energy scale. In addition to the relation between TOF and energy given by the equation, 
\begin{equation}
E=\frac{M}{2}\left(\frac{H^2}{t^2}-gH+\frac{g^2 t^2}{4}\right),
\label{eq:TOFenergyrel}
\end{equation}
it was also necessary to take into account the nonlinear relation between the widths of energy and time channels on the abscissa axis. The relation between $N_{t}$ and $N_{E}$ values, which were proportional to the number of counts in time and energy channels, is given by

\begin{equation}
N_{E}=N_{t}\left[ M\left(\frac{H^2}{t^3}-\frac{g^2 t}{4}\right)\right]^{-1},
\label{eq:Ncountsrel}
\end{equation}
where $M$ is the neutron mass, $g$ is the free fall gravity acceleration, and $H$ is the difference in height between the Fourier modulator and the detector. The latter was 72.5 cm.  

\section{Diffraction grating}
\begin{figure}[H]
	\vspace{3ex}
	\centering
	\minipage[t]{0.43\linewidth}
	\includegraphics[width=\linewidth]{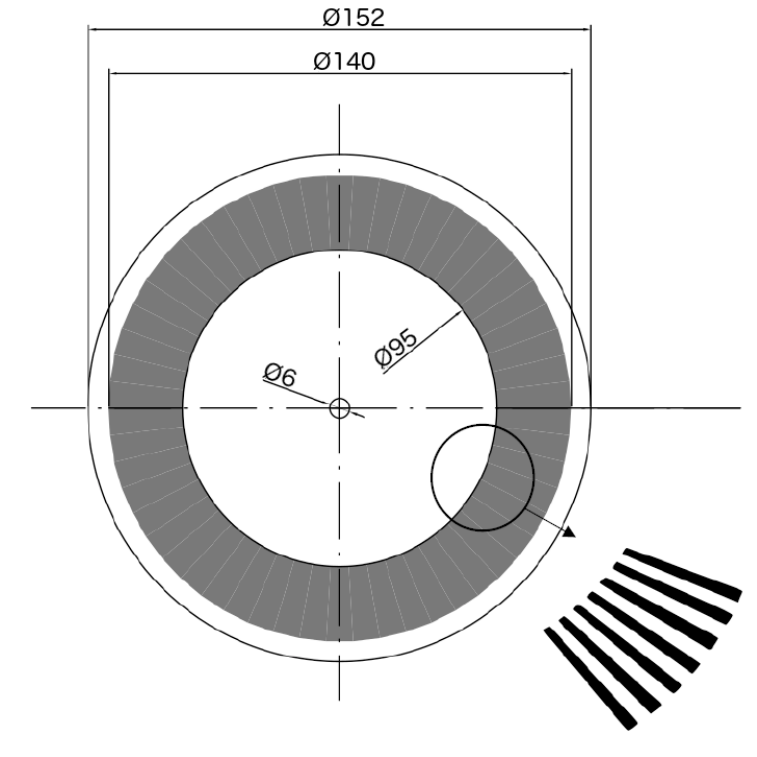}
	\vspace{-8ex}
	\caption{Diffraction grating: dimensions and orientation of grooves.}
	\label{fig:gratscheme}
	\endminipage\hspace{0.02\linewidth}
	\minipage[t]{0.52\linewidth}
	\includegraphics[width=\linewidth]{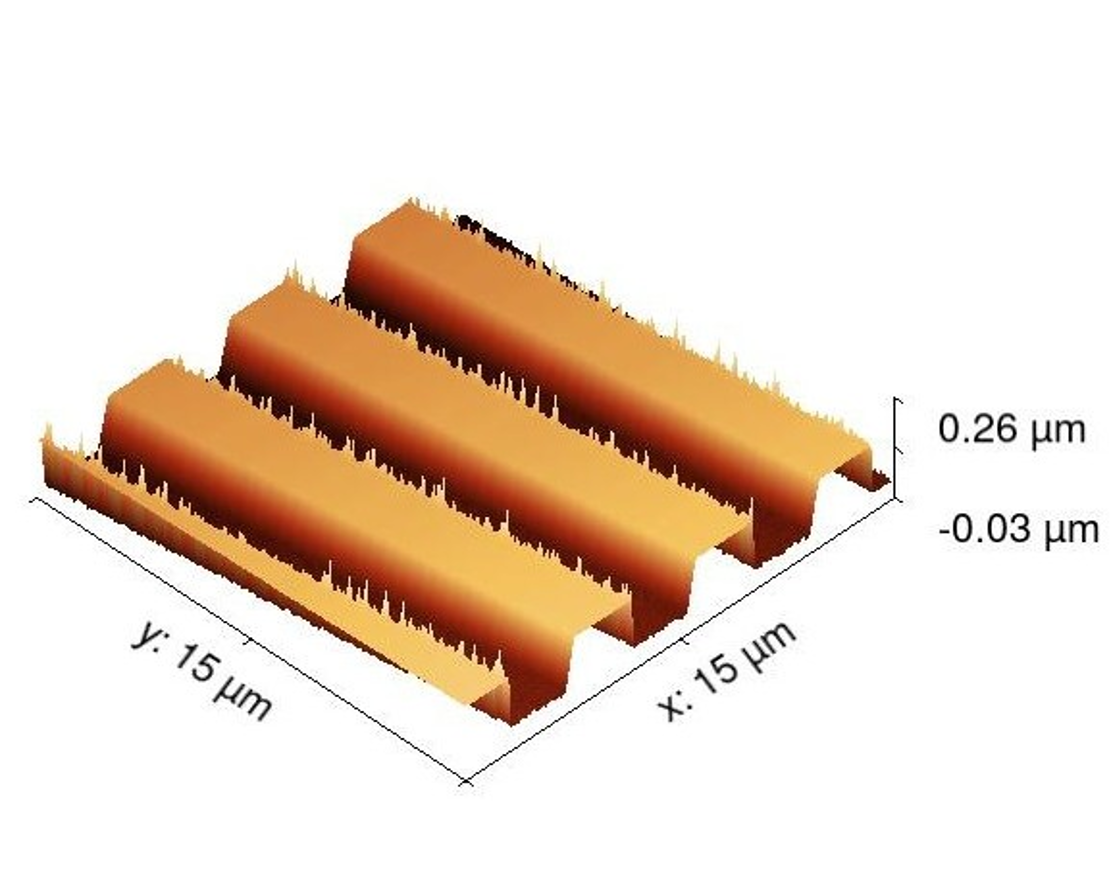}
	\vspace{-8ex}
	\caption{Three-dimensional image of the grating fragment obtained with an atomic force microscope.}
	\vspace{1ex}
	\label{fig:grAFmicrosocope}
	\endminipage
\end{figure}

As in \cite{FrankPhysLettA3112003, FrankJETPLett812005, FrankJETPLett862007, FrankNIMA6112009, KulinNIMA7922015} the diffraction grating was prepared on the surface of a silicon disk 150 mm in diameter and 0.6 mm thick. Radial grooves (see Figs. \ref{fig:grprofile} and \ref{fig:gratscheme}) were made in the peripheral region of the disk, which is a ring with an average diameter of 12 cm and a width of about 2 cm. The widths of the grooves are proportional to the radius and this proportionality ensures a constant angular distance between the grooves equal to a half period. The angular period of the structure is exactly known to be $\alpha = 2pi/N$ with $N = 94500$. The design depth of the grooves of 0.144 nm was chosen to ensure a phase difference $\Delta\phi = \pi$ between the neutron waves passing through the neighboring elements of the grating. The grating was manufactured by Qudos Technology Ltd.\footnote{Qudos Technology Ltd, Rutherford Appleton Laboratory, OX11 0QX Chilton, U.K.}

\begin{figure}[h]
	\vspace{2ex}
	\centering
	\minipage[t]{0.37\linewidth}
	\includegraphics[width=\linewidth]{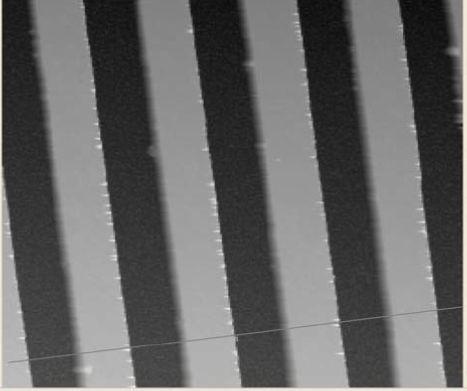}
	\endminipage\hspace{0.08\linewidth}
	\minipage[t]{0.43\linewidth}
	\includegraphics[width=\linewidth]{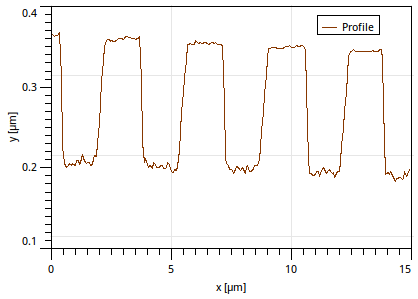}
	\endminipage
	\vspace{-5ex}
	\minipage[t]{0.48\linewidth}
	\caption{Two-dimensional image of the grating fragment and the line showing how the cross section was done for the subsequent analysis.}
	\label{fig:grat2Dimage}
	\endminipage\hspace{0.02\linewidth}
	\minipage[t]{0.46\linewidth}
	\caption{Grating profile measured along the line shown in Fig. \ref{fig:grAFmicrosocope}.}
	\label{fig:grprofilemeas}
	\endminipage
\end{figure}

The grating was examined using an atomic force microscope. Two fragments of the grating positioned at different distances from the center were scanned. The image of one of these fragments is shown in Fig. \ref{fig:grAFmicrosocope}. Each of the two fragments was investigated in four cross sections. The profile in each cross section was plotted for subsequent analysis as that shown in Figs. \ref{fig:grat2Dimage} and \ref{fig:grprofilemeas}. The following parameters were measured: width of the tooth at the top and bottom, width of the groove at the top and bottom, and period of the grating in the planes of the tops and bottoms of the tooth and depth of the groove. The obtained data were averaged. The typical number of averaged measurements was 15--20.

As a result, the following data were obtained. The periods of the grating in the investigated regions were 4.25(1) and 3.40(1) $\mu$m. These data are quite reasonable but do not carry important information because the absolute coordinates of the fragments were not determined. The visible profile of the groove is not exactly rectangular and the groove tapers with depth. This taper is about 50 nm, which corresponds to the expected systematic error of the method. It can be concluded with sufficient confidence that the difference from the rectangular profile, if it really exists, is small. The measured depth of the grooves was 141(2) and 151(2) nm for both fragments, which is in quite satisfactory agreement with the design value of $144 \pm 5$ nm. For the ratio of the tooth-to-groove widths measured on the plane of the tooth tops the value of $\zeta$=0.817(2) was obtained for both fragments.

	
	
\section{Measured energy spectra and comparison with the theory}

Energy spectra of UCN passing through the grating were measured both for the grating 
at rest and for the grating rotating at frequencies of 1500, 3600, and 4800 rpm. With the fixed and specified number of grooves $N$ the expected value of the energy splitting is $\Delta E=2\pi\hbar Nf$ where $f$ is the rotation frequency. Time-of-flight spectra reconstructed in accordance with (\ref{eq:TOFspecseries}) were then recalculated into the energy scale using Eqs. (\ref{eq:TOFenergyrel}) and (\ref{eq:Ncountsrel}).

Theoretically calculated spectra were obtained using the above equations of the dynamical theory of diffraction. The system of Eqs. (\ref{eq:sysreducdiffeqs}) was solved numerically using the Runge-Kutta fourth-order method. To simplify the calculations, it was assumed that the grating has a linear structure rather than a radial one. The following parameters were used for the calculations. The grating period and its linear velocity were obtained as $d = 2\pi R/N$ and $V = 2\pi fR$ cm s$^{-1}$ where $R$ = 6 cm. The height of the tooth was $h$ = 145 nm, and the tooth width-to-period ratio was $\xi$ = 0.45.

The normal component of the neutron velocity at the surface of the grating was taken to be $V_{z}$ = 4.67 m s$^{-1}$ (energy of 114 neV) which corresponds to the position of the maximum of the monochromator transmission spectrum measured earlier by the time-of-flight method. The distribution of the velocity $V_{z}$ was assumed to be a Gaussian with the degree of monochromatization $\Delta V_{z}/V_{z}=0.0175$, where $\Delta V_{z}$ is FWHM of the distribution. The dependence of the resolution on the time of flight was neglected. The poorly known spectrum of horizontal velocities was also assumed to have a Gaussian distribution with a zero average value and a FWHM equal to 3 m s$^{-1}$.

\begin{figure}[H]
	\centering
	\minipage{0.42\linewidth}
	\includegraphics[width=\linewidth, page=1]{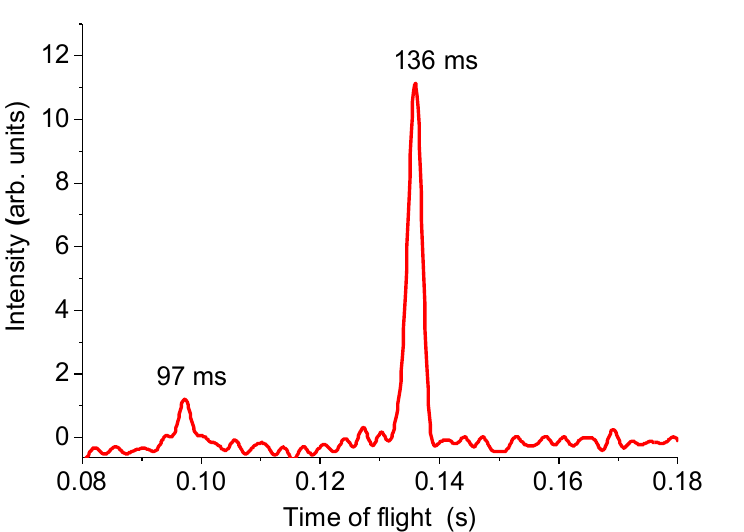}
	\endminipage\hspace{0.05\linewidth}
	\minipage{0.38\linewidth}
	\includegraphics[width=\linewidth, page=2]{figure9}
	\endminipage
	\caption{TOF spectrum (at the left) and energy spectra of UCNs passing through the stationary grating.}
	\label{fig:TOFSpecstatgrat}
\end{figure}

\begin{figure}[H]
	\begin{minipage}[t]{0.4\textwidth}
		\includegraphics[width=\textwidth]{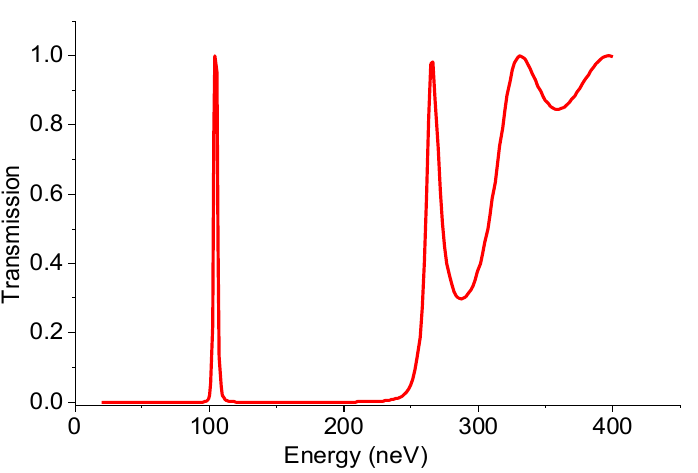}
	\end{minipage}\hspace{0.02\linewidth}
	\begin{minipage}[t]{0.52\textwidth}
		\vspace{-17ex}
		\caption{Calculated transmission spectrum of the filter monochromator. It is known that the real spectrum slightly differs from the calculated one.} \label{fig:calcspec}
	\end{minipage}
\end{figure}

Figure \ref{fig:TOFSpecstatgrat} shows the TOF spectrum of the UCN passing through the stationary grating and its corresponding energy spectrum. In addition to the main peak one can also clearly see a small peak at an energy of 255 neV. Apparently, this is a peak at the energy slightly higher than the value of the potential (see Fig. \ref{fig:calcspec}) significantly suppressed by the transmission of the superwindow.

\begin{figure}[H]
	\centering
	\minipage{0.38\linewidth}
	\includegraphics[width=\linewidth, page=1]{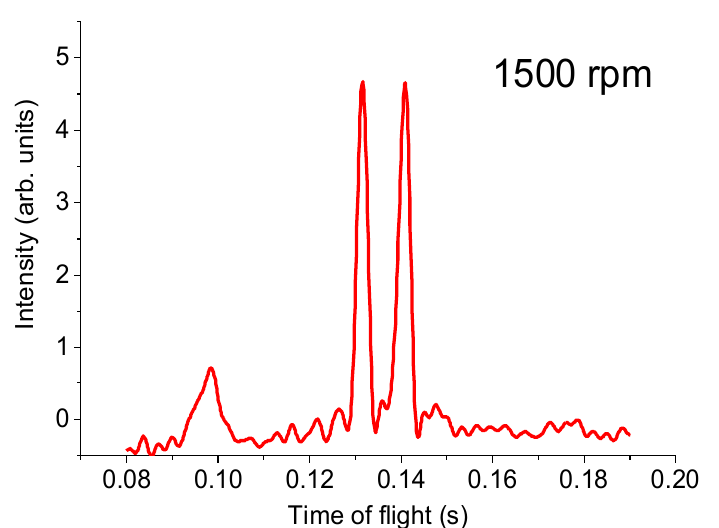}
	\endminipage\hspace{0.05\linewidth}
	\minipage{0.37\linewidth}
	\includegraphics[width=\linewidth, page=2]{figure11}
	\endminipage
	\vspace{0\linewidth}
	\minipage{0.38\linewidth}
	\includegraphics[width=\linewidth, page=3]{figure11}
	\endminipage\hspace{0.05\linewidth}
	\minipage{0.37\linewidth}
	\includegraphics[width=\linewidth, page=4]{figure11}
	\endminipage
	\vspace{0\linewidth}
	\minipage{0.38\linewidth}
	\includegraphics[width=\linewidth, page=5]{figure11}
	\endminipage\hspace{0.05\linewidth}
	\minipage{0.37\linewidth}
	\includegraphics[width=\linewidth, page=6]{figure11}
	\endminipage
	\caption{Time-of-flight (at the left) and energy (at the right) spectra of neutrons diffracted by a moving grating measured at different rotation frequencies. Experimental (red solid line) and theoretical (blue dotted line) spectra are shown.}
	\label{fig:TOFmovinggrat}
\end{figure}

Figure \ref{fig:TOFmovinggrat} illustrates the main results of the experiment. The obtained time and energy UCN spectra measured at three rotation velocities of the grating are displayed together with the results of the calculation. The absolute normalization of the spectra was performed for better visualization purposes.

\section{Conclusion}

As noted in the Introduction the nonstationary splitting of neutron spectra by a moving grating was first observed in \cite{FrankPhysLettA3112003, FrankJETPLett812005}. The results of those first experiments were in good agreement with the theoretical predictions based on the simplest kinematic approach. In this theory only odd diffraction orders may appear in the spectrum and the intensity of ? 1 orders should amount in aggregate to about 80 $\%$ of the total intensity.
The intention to increase the energy splitting of spectra stimulated the use of gratings with a much smaller spatial period. In \cite{KulinJINRCommP32004207} it was supposed that as the ratio of the grating velocity to its period decreases the intensity of the first-order line diminishes and the intensity of even orders (including zero) grows. The qualitative confirmation of these considerations was obtained later and the expected presence of even orders posed serious systematic problems in the experiment \cite{KulinNIMA7922015}. It became evident that the phenomenon of neutron diffraction by a moving grating should be investigated in more detail. The results of \cite{BushuevarXivJETP2016} contributed to a better understanding of the theoretical aspects of the phenomenon. The present work is a new step in its experimental investigation.
UCN spectra appeared in the experiments on neutron diffraction by a moving grating were measured using the time-of-flight Fourier spectrometer. For the first time the diffraction lines of five orders were observed simultaneously. In spite of the fact that the first measurements were made with not very high statistics the results allow a reliable comparison with the theoretical prediction. The obtained data testify that the experiment is in rather satisfactory agreement with the theoretical predictions based on the multiwave dynamical theory of neutron diffraction in the approximation of slowly changing amplitudes \cite{BushuevarXivJETP2016}. 

\begin{acknowledgments}
The authors are very grateful to T. Brenner for his outstanding technical support. This work was partly supported by the Russian Foundation for Basic Research (RFBR Grants No. 15-02-02367 and No. 15-02-02509). 
\end{acknowledgments}

\bibliography{PhysRev.bib}

\end{document}